\documentclass[twocolumn,showpacs,preprintnumbers,amsmath,amssymb]{revtex4}
\usepackage{graphicx}
\usepackage{dcolumn}
\usepackage{bm}
\newcommand{\bef}{\begin{figure}}
\newcommand{\eef}{\end{figure}}

\newcommand{\be}{\begin{equation}}
\newcommand{\ee}{\end{equation}}
\newcommand{\bea}{\begin{eqnarray}}
\newcommand{\eea}{\end{eqnarray}}
\begin{document}

\title{Elliptic flow of thermal dileptons as a probe of QCD matter}

\author{Payal Mohanty$^1$, Victor Roy$^1$, Sabyasachi Ghosh$^1$, 
Santosh K Das$^1$, Bedangadas Mohanty$^2$, 
Sourav Sarkar$^1$, Jan-e Alam$^1$ and Asis K Chaudhuri$^1$}
\medskip
\affiliation{$^1$ Theoretical Physics Division, 
Variable Energy Cyclotron Centre, Kolkata 700064, India}
\affiliation{$^2$ Variable Energy Cyclotron Centre, Kolkata 700064, India}

\date{\today}
\begin{abstract}
We study the variation of elliptic flow of 
thermal dileptons with transverse momentum ($p_T$)  
and invariant mass ($M$) for Pb+Pb collisions at 
$\sqrt{s_{NN}}$  = 2.76 TeV for 30-40\% centrality. 
The dilepton productions from quark gluon plasma (QGP) and hot hadrons
have been considered including the spectral change of light vector 
mesons in the thermal bath. The space time evolution has been 
carried out within the framework of 2+1 dimensional ideal 
hydrodynamics with lattice+hadron resonance gas equation of state. 
We find that a judicious selection of $M$ window can be used to extract 
the collective properties of quark matter, hadronic matter and also get a 
distinct signature of medium effects on vector mesons. 
Results within the ambit of the present work for nuclear collisions 
at Large Hadron Collider (LHC)  
indicate a reduction of elliptic flow ($v_2$)
for $M$ beyond $\phi$ mass, which if observed experimentally 
would imitate the small momentum anisotropy 
in the early stage of the collective motion of the partons.  
We also observe that the magnitude of the elliptic flow 
is significantly larger at LHC than at Relativistic Heavy 
Ion Collider (RHIC) collision conditions. 
\end{abstract}

\maketitle


Collision between nuclei at relativistic energies provides an opportunity
to study Quantum Chromodynamics (QCD) at non-zero temperatures and densities.
Calculations based on lattice QCD (LQCD) predict that at high temperatures
and/or densities hadronic matter melts down to a state 
of deconfined thermal quarks and gluons - a phase of matter 
called quark gluon plasma (QGP). 
The  weakly interacting picture of the QGP
stems from the perception of
asymptotic freedom of QCD at high temperatures and densities.
However, the LQCD calculations indicate
that the deconfined system does not reach the Stefan-Boltzmann
limit corresponding to the non-interacting massless gas of quarks, anti-quarks
and gluons even at a temperature as high as 1000 MeV~\cite{WBlattice}. The RHIC and LHC
are the two experimental facilities where the QGP can be created by colliding
nuclei at high energies. Several probes to study the  
properties of QGP have been proposed.  Among those the 
azimuthal anisotropy or the elliptic flow ($v_2$) of the 
produced particles have been shown to be 
sensitive to the initial condition and the equation of state (EoS) 
of the evolving matter formed in 
heavy ion collision (HIC)(see ~\cite{pasi,teaney,hirano} for review).

Non-central heavy-ion collisions provide an anisotropic
spatial configuration which together with the interactions
among the constituents develop pressure 
gradients of different magnitude along different spatial directions. 
With the expansion the spatial anisotropy reduces and
the momentum space anisotropy 
builds up rapidly. The $v_2$ is a  measure of this momentum space
anisotropy which is defined as:
$v_{2} =\langle\cos(2(\phi-\Psi))\rangle=
\langle{p_x^2-p_y^2}\rangle/\langle{p_x^2+p_y^2}\rangle$,
where $p_x$ and $p_y$ are the $x$ and $y$ component of the particle momenta, $\phi$
is the azimuthal angle of the produced particles and $\Psi$ is the angle subtended by
the plane containing the beam axis and impact parameter with $x$-direction.
Comparison of measured $v_{2}$ with those calculated using relativistic
hydrodynamic and transport approaches have lead to several important results.
The most important of these is the small 
shear viscosity to entropy ratio of the QGP compared to other known fluids~\cite{qm08}.
The mass ordering of $v_{2}$ of identified hadrons,  
clustering of $v_{2}$ separately for baryons and mesons at intermediate $p_{T}$ 
are considered as signatures of partonic coalescence as a mechanism of 
hadron production~\cite{star1,phenix1}. 
In contrast to hadrons, which are predominantly emitted from 
the freeze-out surface of fireball, the electromagnetically 
interacting particles (real photons and lepton pairs) are considered as penetrating
probes ~\cite{mclerran} (see Refs.~\cite{alam1,alam2,rapp}  for review)
which can carry information from the hot interior of the system. 
Therefore, the analysis  of $v_2$ of lepton pairs and photons
can provide  information of the pristine stage of the matter produced in HIC.
The $v_2$ of dileptons~\cite{Rupa} and photons~\cite{dks} 
have been evaluated for RHIC energies and shown that it 
can be used as effective probes to extract the properties
of the partonic plasma. The sensitivity of  $v_2$ of lepton pairs
on EoS has been elaborated in~\cite{Deng} for RHIC collision conditions. 
The lepton pairs are produced from each space time
point of the system and hence the study of $v_2$ of lepton pairs 
will shed light on the time evolution 
of collectivity in the system. 
Further, in addition to $p_T$, dileptons have  an additional
kinematic variable, $M$ unlike photon. 
The evolution of radial flow can be estimated by studying
the $p_T$ spectra of lepton pairs for different $M$ windows
~\cite{renk,payal}.  The radial flow alters the shape 
of the $p_T$ spectra of dileptons - it kicks the low 
$p_T$ pairs to the higher $p_T$ domain, making the 
spectra flatter. Therefore, the presence
of large radial flow may diminish the magnitude of $v_2$ at low $p_T$~\cite{ph}
and this effect will be larger when the radial flow is large i.e. 
in the hadronic phase which corresponds to lepton pairs with 
$M\sim m_\rho$. 

It has been argued that the anisotropic momentum distribution of the hadrons 
can bring the information on the interaction of the
dense phase of the system~\cite{pasi} despite the fact that the hadrons are emitted 
from the freeze-out surfaces when the system is too dilute to support
collectivity. Therefore, a suitable dynamical model is 
required to extrapolate the final hadronic spectra backward in time
to get the information about the early dense phase. 
Such an extrapolation is not required for lepton pairs 
because they are  emitted 
from the entire space-time volume of the system.   
Therefore, the $v_2$ of lepton pairs  provide 
information of the  hot and dense phase directly.
The $v_2$ of dileptons  can also be used to 
to test the validity and efficiency of the extrapolation required for
hadronic $v_2$. 
We will also see below that the $p_T$ integrated 
$M$ distribution of lepton pairs with $M\,(>\,m_\phi$) originate 
from the early time, providing 
information of the partonic phase and pairs with $M\,\lesssim \,m_\rho$ are
chiefly produced later from the hadronic phase. 
Therefore, the $p_T$ integrated $M$ distribution of
lepton pairs may be used as a chronometer of the heavy ion collisions. 
On the other hand, the variation of $v_2$ with $p_T$ for different 
$M$ windows may be used as a flow-meter. 

In this work we study the $M$ and $p_T$ dependence of 
$v_{2}$ for dileptons  using relativistic ideal hydrodynamics~\cite{hydro}
assuming boost invariance along the longitudinal direction~\cite{bjorken} for
LHC collision conditions.
We argue that the $v_2$  of lepton pairs from QGP and hadronic 
phases can be estimated with an appropriate choice of $M$.
In addition to this we also show that 
the thermal effects on the $\rho$ spectral function 
is visible through $v_{2}$ in the low mass (below $\rho$ peak) 
dileptons. 

The elliptic flow of dilepton, $v_2(p_T,M)$ can be defined as:
\begin{eqnarray}
&&v_2=
\frac{\sum\int cos(2\phi)
\left(\frac{dN}{d^2p_TdM^2dy}\arrowvert_{y=0}\right) d\phi}
{\sum\int\left(\frac{dN}{d^2p_TdM^2dy}\arrowvert_{y=0}\right)d\phi }
 \label{eq1}
\end{eqnarray}
where the $\sum$ stands for summation over 
Quark Matter(QM) and Hadronic Matter(HM) phases. 
The quantity $dN/d^2p_TdM^2dy\arrowvert_{y=0}$
appearing in Eq.~\ref{eq1}
can be obtained from the dilepton production 
per unit four volume, $dN/d^4pd^4x$
in a thermalized medium by integrating over the
space-time evolution of the system.  The quantity, 
$dN/d^4pd^4x$ is given by ~\cite{mclerran}:
\begin{equation}
\frac{dN}{d^4pd^4x}=-\frac{\alpha^2}{6\pi^3}\frac{1}{M^2}L(M^2)
\exp\left(-\frac{p_0}{T}\right)g^{\mu\nu}W_{\mu\nu}(p_0,\vec{p})
\label{eq2}
\end{equation}
where $W_{\mu\nu}(p_0,\vec{p})$ is the electromagnetic (EM) current correlator, 
$g^{\mu\nu}$ is the metric tensor, $\alpha$ is the electromagnetic coupling,
$p=(p_0,\vec{p})$ is  the four-momentum of the pair,
$T$ is the temperature of the thermal bath,  
$L(M^2)=\left(1+2m_l^2/M^2\right)\sqrt{1-4m_l^2/M^2}$ 
arises from the Dirac spinors (lepton pair) in the final state   
,$M$ is the invariant mass of the lepton pair 
and $m_l$ is the lepton mass. 

For QGP Eq.~\ref{eq2} leads to the standard rate of lepton pair 
production (in the leading order) from annihilation of $q\bar{q}$ pairs~\cite{cleymans}. 
For the dilepton production from hot hadrons we briefly outline
the dilepton production processes here  and refer to ~\cite{sabya} for details. 
For the low mass dilepton production from HM the decays
of thermal light vector mesons namely $\rho$, $\omega$ and $\phi$ 
have been considered. The change of spectral function of $\rho$
due to its interaction with $\pi,\omega, a_1, h_1$ (see~\cite{sabya,sabyaEPJC}
for details) and baryons ~\cite{eletsky}  have been
included in evaluating the production of lepton pairs from HM. 
For the $\omega$ spectral function the width at non-zero temperature
is taken from Ref~\cite{Weise} and  medium effects  
on $\phi$ is  ignored here. 
The continuum part of the spectral 
function of $\rho$ and $\omega$ have also been included in dilepton 
production rate ~\cite{alam2,cont}. In the present work dileptons from 
non-thermal
sources {\it e.g.} from the Drell-Yan process and decays of heavy
flavours ~\cite{lhcrapp} 
have been ignored in evaluating the elliptic flow of lepton pairs
from QM and HM. If the charm and bottom quarks do not
thermalize then they are not part of the flowing QGP and hence do not
contribute to the elliptic flow.
The model employed in the present work
leads to a good agreement with NA60 dilepton data~\cite{NA60} 
for SPS collision conditions~\cite{jajNA60}.

To evaluate $v_2$ from Eq.~\ref{eq1} one needs to 
integrate the production rate given by Eq.~\ref{eq2} 
over the space time evolution of the system  - from the initial
QGP phase to the final hadronic freeze-out state through a phase transition
in the intermediate stage.  
We assume that the matter is formed in QGP phase 
with negligible net baryon density. 
The  initial condition required to solve the hydrodynamic equations 
for the description of the matter produced 
in Pb+Pb collision at $\sqrt{s_{\mathrm NN}}=2.76$ TeV for 30-40\% centrality 
are taken as follows:
$T_i=456$ MeV is the value of the initial temperature
corresponding to the maximum of the initial energy density  profile at
the thermalization time $\tau_i=0.6$ fm/c. 
The value of the transition temperature,
$T_c$ for quark hadron conversion is taken as 175 MeV. 
The EoS  required to close the hydrodynamic equations  is constructed 
by complementing Wuppertal-Budapest lattice simulation 
~\cite{WBlattice} with a hadron resonance gas comprising of
all the hadronic resonances up to mass of $2.5$ GeV ~\cite{victor,bmja}. 
The energy of the lepton pair ($p_0$) originating from a hydrodynamically
expanding system should be replaced
by its value ($p\cdot u$) in the co-moving frame which is given by:
$p\cdot u=\gamma_T(M_T \cosh(y-\eta)-v_xp_T\cos\phi-v_yp_T\sin\phi)$,
where $u=(\gamma,\gamma\vec{v})$, is the fluid four-velocity, 
$y$ is the rapidity and $\eta$ is the space-time rapidity,
$\gamma_T=(1-v_T^2)^{-1/2}$, $v_T^2=v_x^2+v_y^2$, $v_x$ and $v_y$ 
are the $x$ and $y$ component of the velocity. 
The system is assumed to get out of chemical equilibrium at
$T=T_{ch}=170$ MeV~\cite{tsuda}.  
The kinetic freeze-out temperature $T_{F}=130$ MeV 
is fixed from the $p_T$ spectra of the produced hadrons 
at the same collision energy of Pb+Pb system.
The EoS and the values of the parameters 
mentioned above are constrained by the $p_T$ 
spectra (for $0-5\%$ centrality) and 
elliptic flow (for $10-50\%$ centrality) of charged hadrons~\cite{victor}
measured by ALICE collaboration ~\cite{ALICE}. 

\%
\bef[h]
\begin{center}
\includegraphics[scale=0.25]{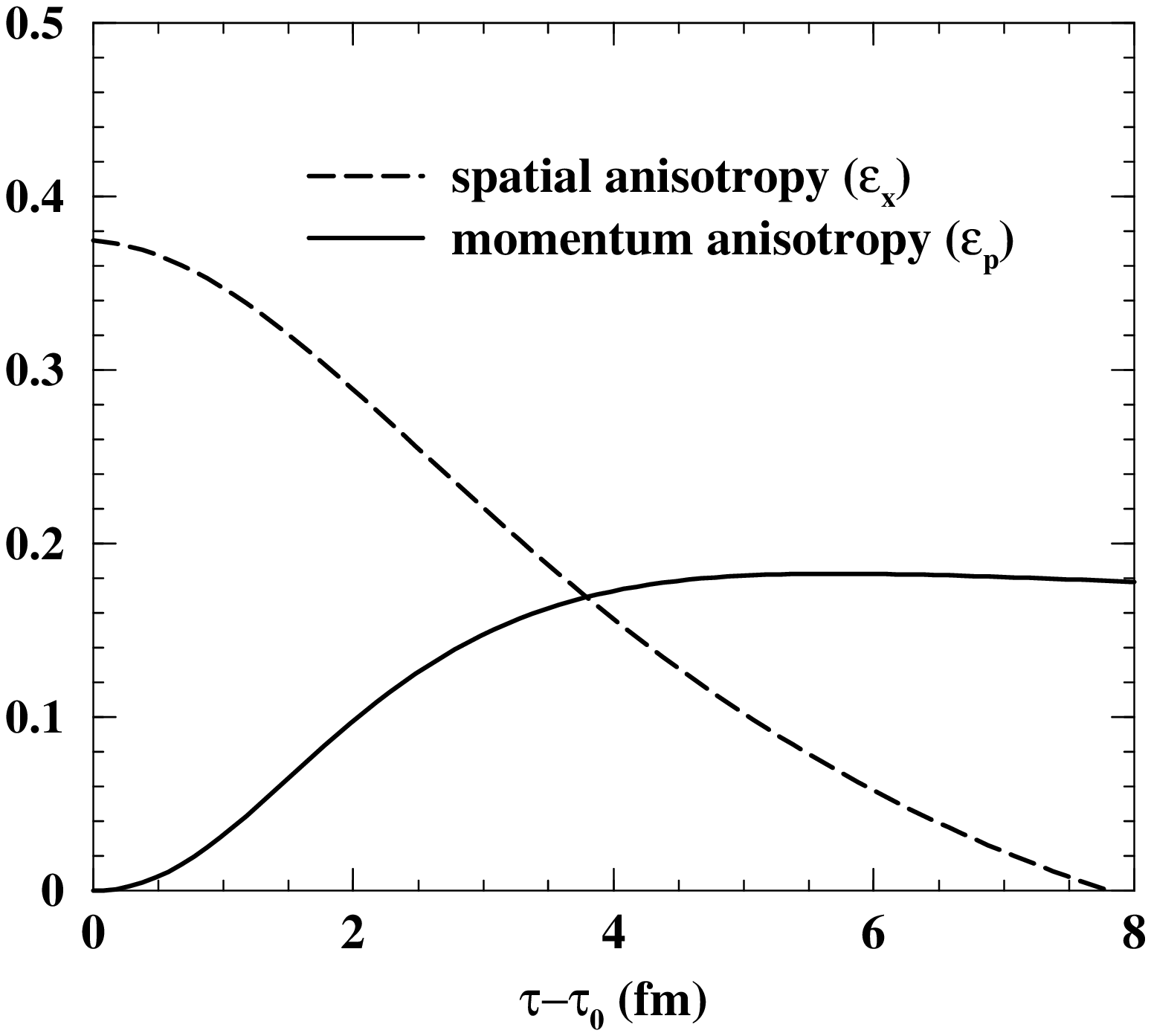}
\includegraphics[scale=0.25]{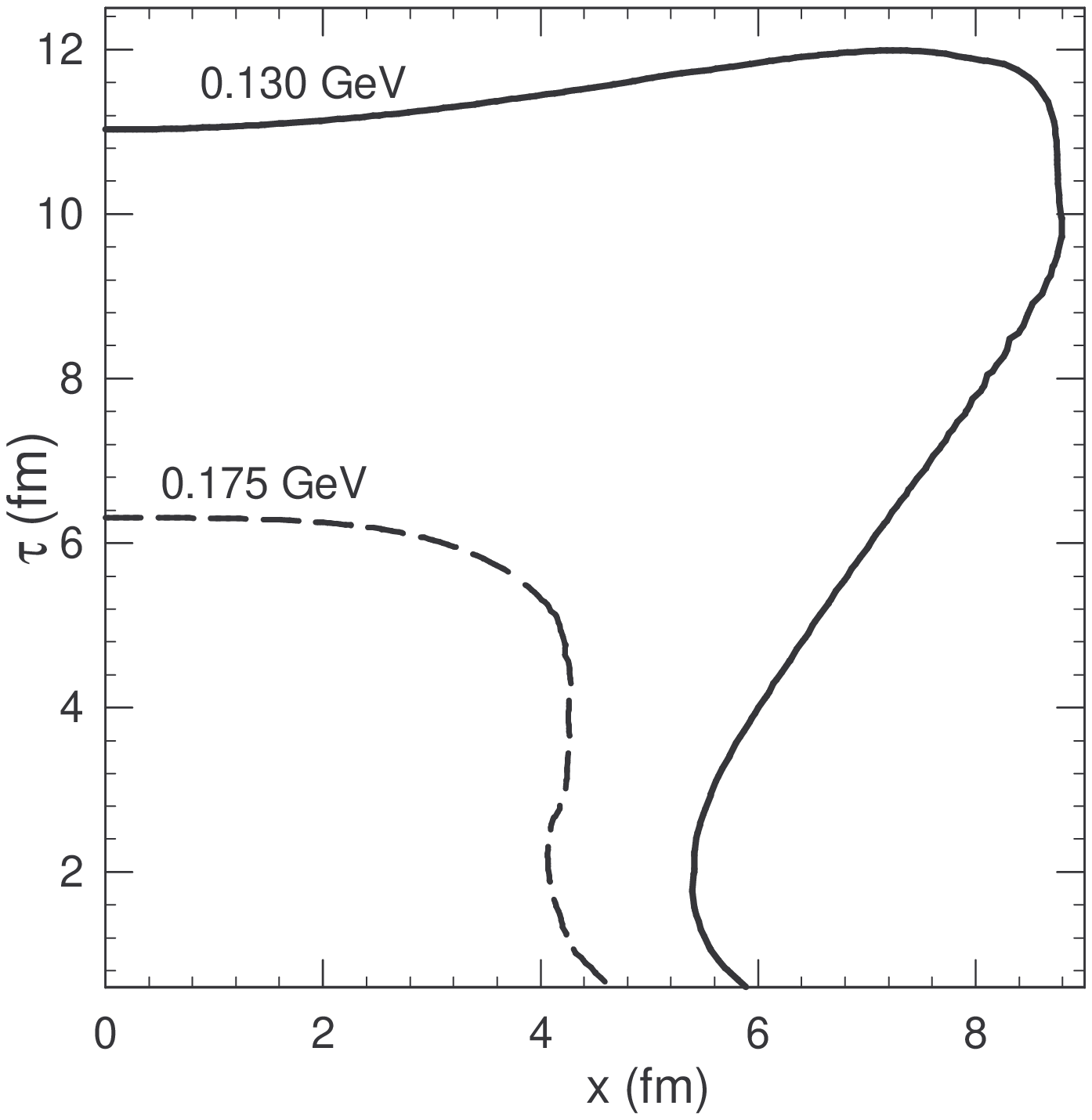}
\caption{Left panel: 
Variations of spatial and momentum space anisotropy 
with proper time. Right panel:
Constant temperature contours denoting the space time boundaries of
the QGP and hadronic phases.  
}
\label{fig1}
\end{center}
\eef
\bef[h]
\begin{center}
\includegraphics[scale=0.22]{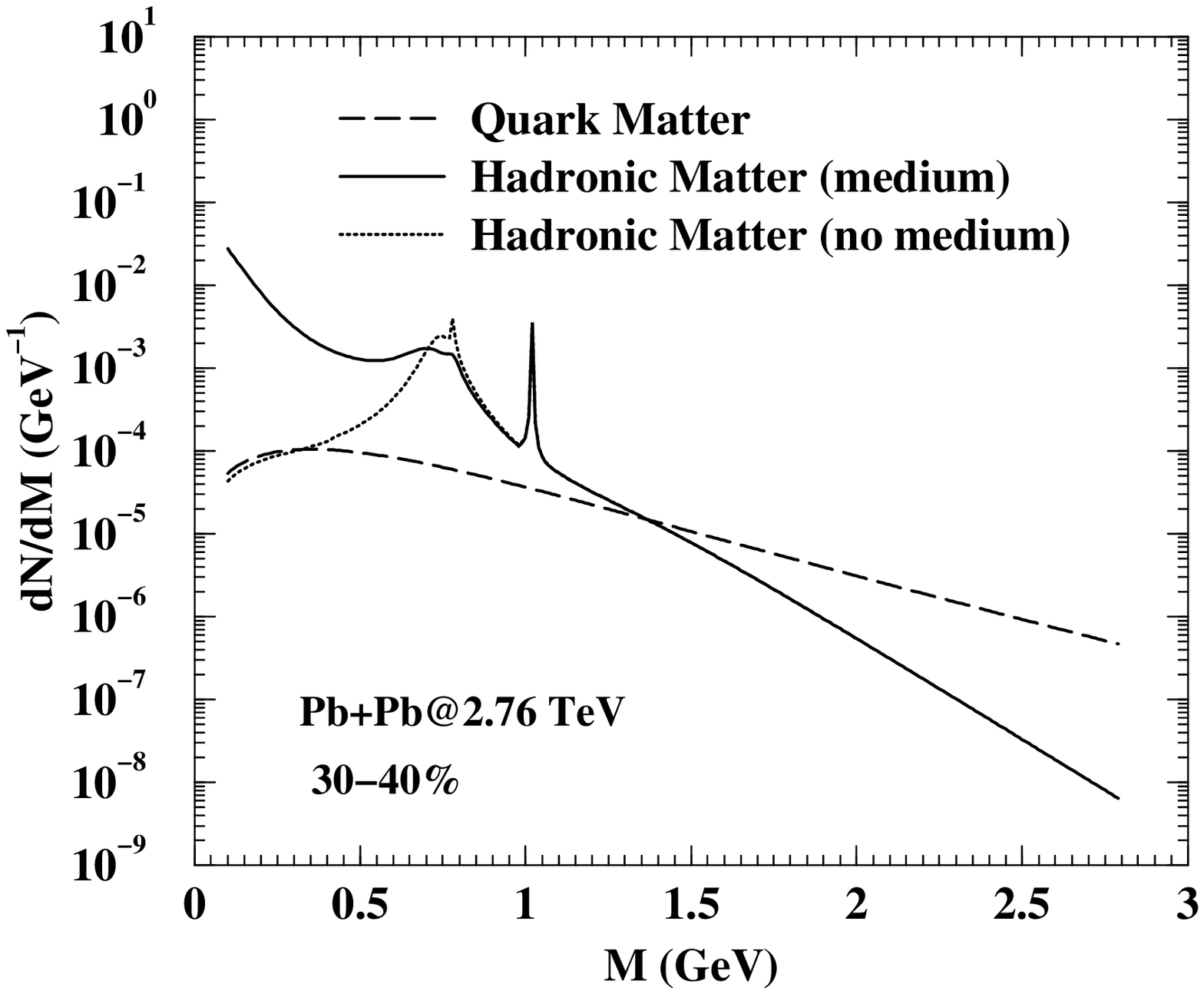}
\includegraphics[scale=0.22]{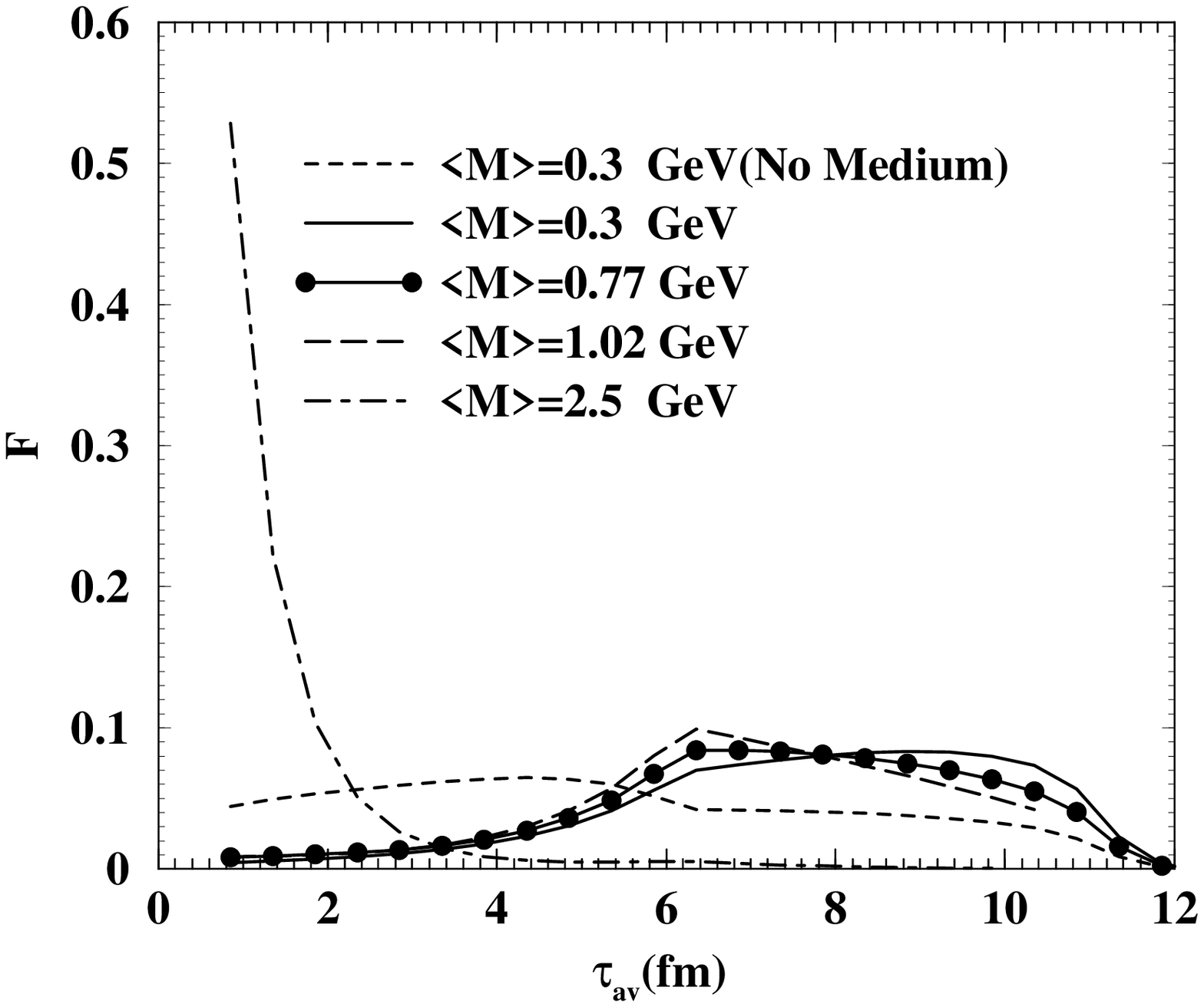}
\caption{Left panel: Invariant mass distribution 
of lepton pairs from quark matter
and hadronic matter (with and without medium effects).
Right panel: Fractional contribution of lepton pairs for various
invariant mass windows as a function of average proper time 
(see text for details). 
}
\label{fig2}
\end{center}
\eef

The spatial anisotropy of the matter produced in non-central HIC is defined as:
$\epsilon_x=\frac{\langle x^2-y^2\rangle}{\langle x^2+y^2\rangle}$,
where the average is taken over the energy distribution of the system formed
in a collision of given impact parameter.
The resulting momentum anisotropy of the interacting system 
produced in non-central HIC is defined as:
$\epsilon_p=\frac{\int dx\int dy [T^{yy}-T^{xx}]}{\int dx\int dy [T^{yy}+T^{xx}]}$,
The variations of $(\epsilon_x)$ and $(\epsilon_p)$ with
time have been displayed in the left panel of 
Fig.~\ref{fig1} for 30-40\% centrality. 
We find that $\epsilon_x$ 
($\epsilon_p$) reduces (increases) with 
time due to the prevailing pressure gradients which 
causes the matter to flow.  Since $v_2\propto\epsilon_p$
therefore, the variations of $v_2$ and $\epsilon$ with $\tau$ are expected to
be similar. In Fig.~\ref{fig1} (right) we depict the constant temperature
contours corresponding to $T_c=175$ MeV and $T_f=130$ MeV in the
$\tau-x$ plane (at zero abscissa) indicating the boundaries for the QM and
HM phases respectively. The life time of the QM phase $\sim 6$ fm/c 
and the duration of the HM is$\sim 6-12$ fm/c.  Throughout this
work by early and late will approximately mean the duration of
the QM and HM respectively.  

With all the ingredients mentioned above we
evaluate the $p_T$ integrated $M$ distribution of lepton pairs
originating from QM and HM (with and
without medium effects on the spectral functions 
of $\rho$ and $\omega$). 
The results are displayed in Fig.~\ref{fig2} (left) for the
initial conditions and centrality mentioned above. 
We observe that for $M\,>\,M_{\phi}$
the QM contributions dominate. For  $M_{\rho}\lesssim M\lesssim M_{\phi}$
the  HM shines brighter than QM. For $M\,<M_{\rho}$, the HM (solid 
line) over shines
the QM due to the enhanced contributions primarily from the  medium
induced broadening of $\rho$ spectral function. However, the contributions
from QM and HM become comparable in this region of $M$ if the medium
effects on $\rho$ spectral function is ignored (dotted line). 
Therefore, the results depicted in Fig.~\ref{fig2} 
(left) indicate that a suitable choice of $M$ window will
enable us to unravel the contributions from a particular phase
(QM or HM). 

To further quantify these issues we evaluate the following quantity:
\begin{eqnarray}
&&F=
\frac{\int^\prime  
\left(\frac{dN}{d^4xd^2p_TdM^2dy}\right)dxdyd\eta\tau d\tau d^2p_TdM^2}
{\int\left(\frac{dN}{d^4xd^2p_TdM^2dy}\right)dxdyd\eta\tau d\tau d^2p_TdM^2}
 \label{eq3}
\end{eqnarray}
where the $M$ integration in both the numerator and 
denominator are performed  for selective  $M$
windows from $M_1$ to $M_2$ with  mean $M$
defined as $\langle M\rangle = (M_1+M_2)/2$. 
The prime in $\int^\prime$ in the numerator indicates 
that the $\tau$ integration
in the numerator is done from $\tau_1=\tau_i$ to $\tau_2=\tau_i+\Delta\tau$ 
with progressive increment of $\Delta\tau$, 
while in the denominator 
the integration is done over the entire lifetime
of the system. In the right panel of Fig.~\ref{fig2},  
$F$ is plotted against $\tau_{\mathrm av} (=(\tau_1+\tau_2)/2)$.
The results substantiate the fact that pairs with high $\langle M\rangle\sim 2.5$ GeV 
originate from QM ($\tau_{\mathrm av}\lesssim 6$ fm/c, QGP phase) 
and pairs with 
$\langle M\rangle\sim 0.77$ GeV mostly emanate from the HM
phase ($\tau_{\mathrm av}\geq 6$ fm/c).  
The change in the properties of $\rho$ due to its interaction
with thermal hadrons in the bath 
is also visible through $F$ evaluated for $\langle M\rangle\sim 0.3$ GeV 
with and without medium effects. 
This clearly indicates that  
the $\langle M\rangle$ distribution of lepton pairs can be exploited 
to extract collectivity  of different phases of the evolving  
matter.

\bef[h]
\begin{center}
\includegraphics[scale=0.4]{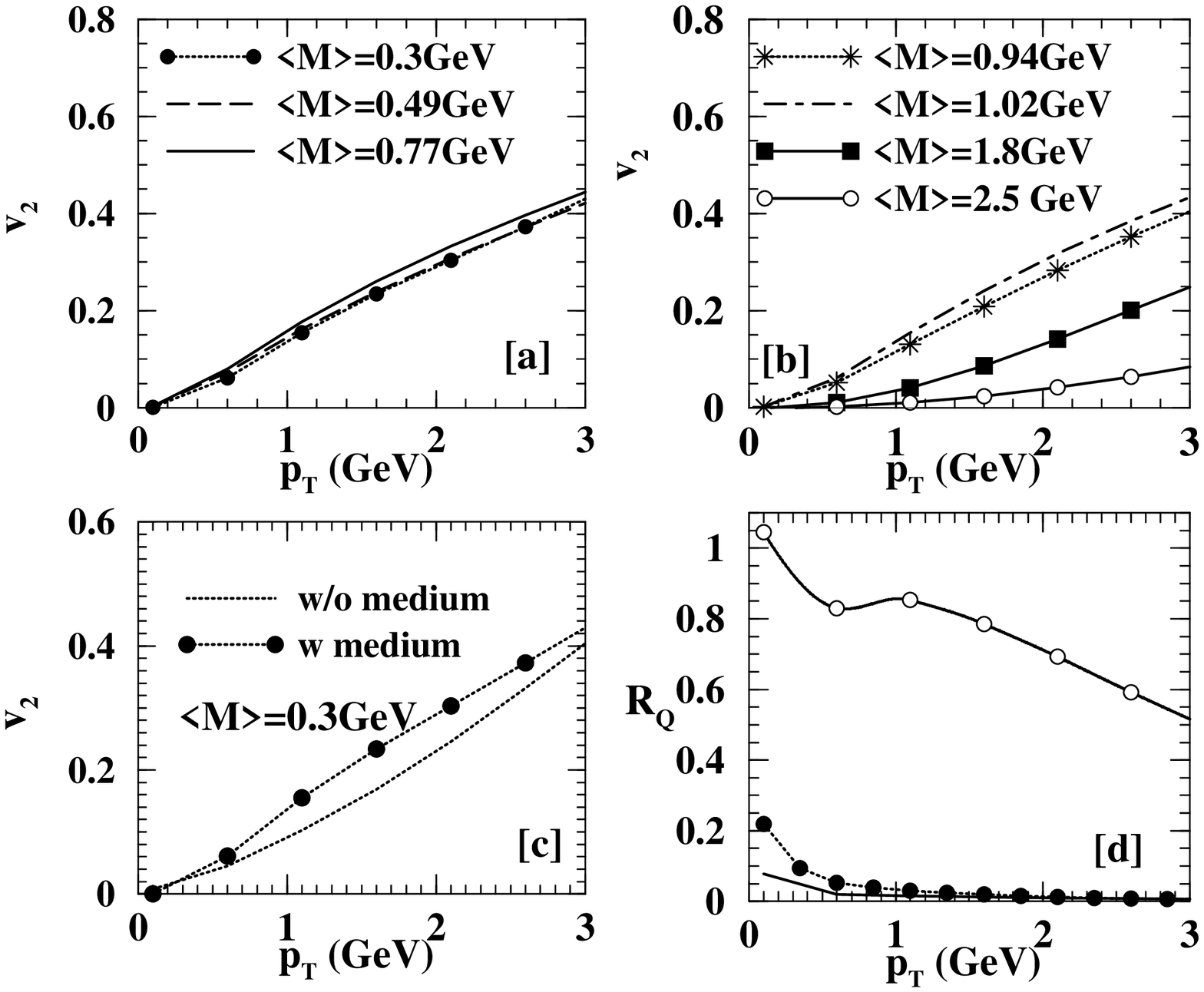}
\caption{[a]  and [b] indicate elliptic flow of lepton pairs 
as a function of $p_T$ for various $M$ windows.  [c] displays the effect 
of the broadening of $\rho$ spectral function on the elliptic flow
for $\langle M\rangle = 300$ MeV.  
[d] shows the variation of $R_Q$  (see text) 
with $p_T$ for $\langle M\rangle=0.3$ GeV (solid circle), 0.77 GeV (line)
and 2.5 GeV (open circle). All the results displayed here are for 30-40\% centrality.
}
\label{fig3}
\end{center}
\eef

\par
Fig.~\ref{fig3} ([a] and [b]) show the differential elliptic 
flow, $v_2(p_T)$ of dileptons arising from various $\langle M\rangle$
domains. We observe that for  $\langle M\rangle=2.5$ GeV, $v_2$ is 
small for the entire $p_T$ range 
because these pairs arise dominantly from the QM epoch (see 
Fig.~\ref{fig2}, right panel) when the flow is not developed fully.  
By the time (6-12 fm/c) when the pairs are emitted predominantly 
in the region $\langle M\rangle=0.77$ GeV,
the flow is  fully developed which gives rise to large $v_2$. 
It is also interesting to note that the medium induced enhancement 
of $\rho$ spectral function provides a visible modification 
in $v_2$ for dileptons below $\rho$ peak (Fig.~\ref{fig3} [c]). 
The medium-induced effects lead to an enhancement of $v_2$  
of lepton pairs which is culminating from the `extra' interaction
(absent when a vacuum $\rho$ is considered) of 
the $\rho$ with other thermal hadrons in the bath.
We note that the differential elliptic
flow, $v_2(p_T)$ obtained here at LHC is larger than
the values obtained  at RHIC~\cite{Rupa,Deng} for
all the invariant mass windows. 
In Fig.~\ref{fig3} [d] we depict the variation of  $R_Q$ with $p_T$ for
$\langle M\rangle=0.3$ GeV (line with solid circle) 0.77 GeV (solid line)
and 2.5 GeV (line with open circle). The quantity $R_Q$ ($R_H$) is defined as,
$R_Q=v_2^{\mathrm QM}/ 
(v_2^{\mathrm QM}+v_2^{\mathrm HM})$ 
[$R_H=v_2^{\mathrm HM}/ 
(v_2^{\mathrm QM}+v_2^{\mathrm HM})$] 
where $v_2^{\mathrm QM}$ and $v_2^{\mathrm HM}$ are the elliptic flow of QM and 
HM respectively. 
The results clearly illustrate 
that $v_2$ of lepton pairs in the large $\langle M\rangle(=2.5$ GeV) domain 
(open circle in Fig.~\ref{fig3} [d]) originate from QM for 
the entire $p_T$ range considered here.  The value of $R_Q$ is large in this domain
because of the large (negligibly small) contributions from QM (HM) phase.
It is also clear that the contribution
from QM phase to the elliptic flow for  
$\langle M\rangle(=0.77$ GeV)  is very 
small (solid line in Fig.~\ref{fig3} [d]). The value of
$R_H$ for $\langle M\rangle =0.77$ GeV is large (not shown in the
figure). 

\bef
\begin{center}
\includegraphics[scale=0.3]{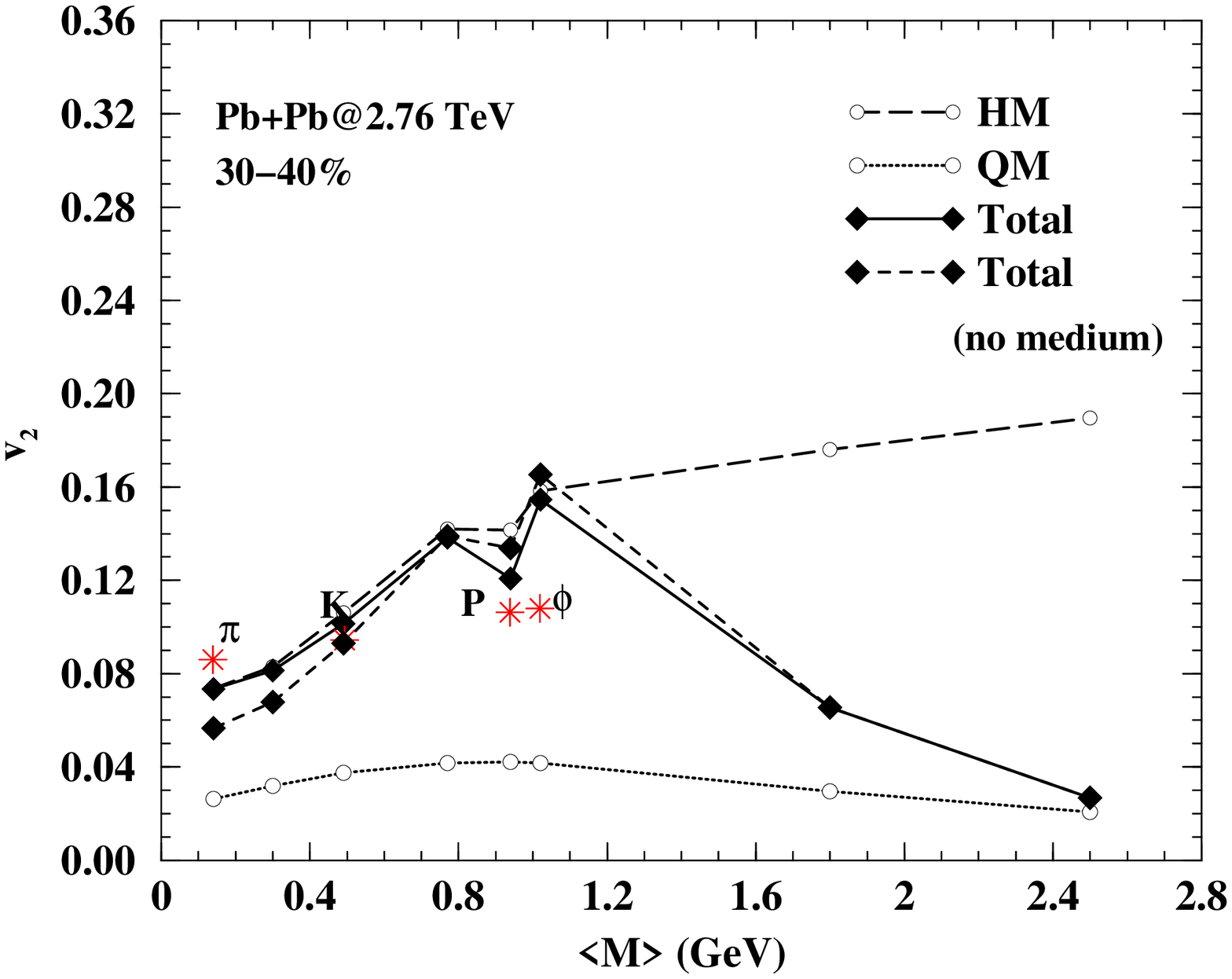}
\caption{(Color online) Variation of dilepton elliptic flow as 
function of $\langle M\rangle$ for QM, HM (with and without medium effects) and 
for the entire evolution. The symbol $*$ indicates the value of $v_2$ for hadrons
{\it e.g.} $\pi$, kaon, proton and $\phi$.
}
\label{fig4}
\end{center}
\eef

The $v_2$ at the HM phase (either at $\rho$ or
$\phi$ peak) is larger than its value 
in the QGP phase (at $\langle M\rangle=2.5$ GeV,
say) for the entire $p_T$ range considered here. 
Therefore, the
$p_T$ integrated values of $v_2$ should also retain this character 
at the corresponding values of $\langle M\rangle$, which is clearly observed
in Fig.~\ref{fig4} which displays  the variation of
$v_2(\langle M\rangle)$  with $\langle M\rangle$. 
The $v_2$ ($\propto\epsilon_p$) of QM is small because
of the small pressure gradient in the QGP phase. 
The $v_2$ resulting from hadronic phase has a peak around $\rho$ pole indicating
the full development of the flow in the HM phase. 
For $\langle M\rangle\>>\, m_\phi$ the $v_2$ obtained from the 
combined phases approach the value corresponding to the $v_2$
for QGP.  Therefore, measurement of $v_2$ for large $\langle M\rangle$
will bring information of the QGP phase at the earliest time of the evolution. 
It is important to note that the $p_T$ integrated $v_2(\langle M\rangle)$
of lepton pairs with $\langle M\rangle\,\sim m_\pi, m_K$
is close to
the hadronic $v_2^\pi$ and $v_2^K$ (symbol $*$ in Fig.~\ref{fig4}) 
if the thermal effects on $\rho$
properties are included. Exclusion of medium effects give lower
$v_2$ for lepton pairs compared to hadrons. 
The fact that the $v_2$ of the (penetrating) lepton pairs are similar 
in magnitude to the $v_2$ of hadrons for ($\langle M\rangle\sim m_\pi, m_K$, 
$m_{\mathrm proton}$ etc), it ascertains
that the anisotropic momentum distribution of hadrons carry
the information of the HM phase with duration $\sim 6-12$ fm/c
(left panel of Fig.~\ref{fig1}).   
We also observe that the variation of $v_2(\langle M\rangle)$ 
with $\langle M\rangle$ has a structure similar to $dN/dM$ vs $M$. 
As indicated by Eq.~\ref{eq1} we can write
$v_2(\langle M\rangle)\sim \sum_{i=QM,HM} v_2^{\mathrm i}\times f_{\mathrm i}$,
where $f_i$ is the fraction of QM or HM from various space-time regions.
The structure of $dN/dM$ is reflected in $v_2(\langle M\rangle)$ through $f_i$. 
We find that the magnitude of $v_2(\langle M\rangle)$ at LHC is larger 
than its value at RHIC.

In conclusion, we have evaluated the $v_2$ of  dileptons
originating from the Pb+Pb collisions at $\sqrt{s_{NN}}$ = 2.76 TeV for $30-40\%$
centrality. Our study shows that $v_{2}(M)$ provides useful information
on the collective motion of the evolving QCD matter formed in high energy heavy-ion collisions.
If the heavy quarks (charm and bottom) produced in HIC do not thermalize and hence not 
become part of the flowing QGP  then the lepton pairs which originate from the decays of 
heavy flavours will not contribute to the $v_2$ of lepton pairs. 
In such a scenario an experimental observation of the reduction of $v_{2}(M)$ with 
increasing $M$ beyond $\phi$ mass 
would reflect the presence of small momentum space 
anisotropy through modest collective motion in the QM phase. 
We observe that $v_2(\langle M\rangle)$ of the penetrating 
probe (lepton pairs) for $\langle M\rangle=m_\pi$ and $m_K$ is similar
to the hadronic $v_2^\pi$ and $v_2^K$ when the medium induced change in the $\rho$ spectral
function is included in evaluating the dilepton spectra. 
The medium effects are large during the dense phase of the hadronic system, therefore, 
this validates the findings that the hadronic $v_2$ carry the information
of the dense part of the hadronic phase.
Our study also establishes the fact that the invariant mass 
dependence of dilepton $v_{2}$ can in principle act as a clock for the space
time evolution of the system formed in HIC.


{\bf Acknowledgment:}  
PM, SKD and JA are partially supported by DAE-BRNS project Sanction No.
2005/21/5-BRNS/2455. BM is partially supported by DAE-BRNS
project Sanction No. 2010/21/15-BRNS/2026.

\end{document}